\documentclass{article}

     \PassOptionsToPackage{numbers, compress}{natbib}

\usepackage[final]{neurips_2019}

\usepackage[utf8]{inputenc} 
\usepackage[T1]{fontenc}    
\usepackage{hyperref}       
\usepackage{url}            
\usepackage{booktabs}       
\usepackage{amsfonts}       
\usepackage{nicefrac}       
\usepackage{microtype}      
\usepackage{color}
\usepackage{graphicx}
\usepackage{epstopdf}
\usepackage{caption}
\usepackage{hyperref}
\hypersetup{
  colorlinks=true,
    linkcolor=blue,
    filecolor=magenta,      
    urlcolor=cyan,
}

\title{Online tuning and light source control using a physics-informed Gaussian process}

\author{%
 A. Hanuka\thanks{adiha@slac.stanford.edu} \hfill  {, J. Duris, }  \hfill {J. Shtalenkova, } \hfill {D. Kennedy, } \hfill  {A. Edelen, }\hfill
 \hfill {D. Ratner, } \hfill
  {X. Huang}  \\ \\
  SLAC National Accelerator Laboratory\\ Menlo Park\\ CA 94025, USA \\
  \texttt{} \\
}

\begin{document}

\maketitle

\begin{abstract}
  Operating large-scale scientific facilities often requires fast tuning and robust control in a high dimensional space. In this paper we introduce a new physics-informed optimization algorithm based on Gaussian process regression.
Our method takes advantage of the existing domain knowledge in the form of realizations of a physics model of the observed system. 
  We have applied a physics-informed Gaussian Process method experimentally at the SPEAR3 storage ring to demonstrate online accelerator optimization. This method outperforms Gaussian Process trained on data as well as the standard approach routinely used for operation, in terms of convergence speed and optimal point.
  The proposed method could be applicable to automatic tuning and control of other complex systems, without a prerequisite for any observed data.
  
\end{abstract}

\section{Introduction}

Machine learning tuning algorithms have the ability to learn from archived data and incorporate physical models in order to improve both exploration and exploitation. 
Historical data may be insufficient to model a complex system prone to drifting variables, measurement errors, and non-uniform sampling of parameter space.
Physical theories or simulations of an experiment are often far less costly to evaluate and can capture the qualitative dynamics better than archive data.
Therefore, incorporating physics models may increase the speed of convergence and robustness of an online tuning process. 


In this work we present a physics-informed approach using Bayesian optimization using Gaussian processes for online tuning of an objective. While a data-driven approach would obviously fail for new machines and configurations, the physics-informed approach requires no data, and is more robust and flexible.
We set a Bayesian optimization routine to control a system in an unknown state possibly far from its optimum.
The Bayesian optimizer uses two components: 1) an online statistical model of how the objective responds to controls; 
and 2) an acquisition function or control policy which decides where to look based on the current state of the model built from the observed data.


The statistical model we chose is a Gaussian Process (GP), which balances between data fit and the complexity of the model \cite{rw:gpml_book}. 
The Gaussian process learns patterns in data by employing a kernel, or covariance function, to describe similarities between acquired points. 
The selection of kernels is one of the critical steps in achieving an operational Gaussian process. There is a need to develop more sensible kernel functions which allow for the incorporation of prior physics knowledge and help us to gain real insight into the system.

In what follows we estimate the kernel parameters in two ways: first, using Bayesian inference on historical data, and then, using an intuitive approach of estimating basis functions from a physical model. Finally, we experimentally demonstrate that the second approach to estimate kernel parameters increases the speed and effectiveness of the optimization.



\section{Gaussian Process\label{sec:GP}}
A Gaussian process (GP) is a non-parametric model which calculates probability densities over a space of functions, providing a practical, probabilistic approach to kernel machine learning \cite{rw:gpml_book}. GP model is attractive in many fields, since it can assemble a lot of data complex structures in it.
Whereas a Gaussian distribution is characterized by a normal distribution with mean and covariance $y \sim {\sc N} ({\mu},\Sigma)$, a Gaussian \textit{process} is a normal distribution over mean and covariance \textit{functions} $f(x) \sim {\sc GP} (m(x),k(x,x'))$. 
This covariance function (kernel) could be defined as a convolution of a basis function $\phi(x)$ with itself \cite{w:gpml_book_ch4}:
\begin{equation}
    k({x_i},{x_j}) \propto \int_{-\infty}^{\infty} \phi({x_i} - {c}) \phi({x_j} - {c}) d{c}
    \label{eq:kernel_from_basis}
\end{equation}
where c denotes the mean of the basis function. For example, Gaussian-shaped basis functions of the form $\phi({x})=\exp(-({x}-{c})^{T}\Sigma({x}-{c}))$ are useful for modeling many smooth functions. 
If there are no correlations between devices, $\Sigma={\rm diag}(l)^{-2}$ where $l$ is a vector of characteristic length-scales.








\subsection{Data-informed kernel}

Kernels and their parameters are chosen by maximizing the marginal likelihood of the GP regression given historical training data \cite{rw:gpml_book}.
In order to incorporate high-level structure into the models, new kernel functions with different properties can be generated by using simpler kernels as building blocks \cite{duvenaud:kernel_composition,Wilson2013,Bozson2018, Frate2017}, by applying a nonlinear transform to the input data \cite{DKL, manifoldgp}, or by cascading GPs \cite{Deep-GP}. 

In cases of sparse sampling of high dimensional archived data where correlations between features are difficult to resolve, it can be advantageous to combine correlations from simulation with length-scale fit from data \cite{Duris2019}. Calculating a kernel from data becomes impractical when preparing for new experiments where relevant data does not exist. One approach is to select a kernel which maximizes the marginal likelihood with {\it simulated} rather than experimental data \cite{Yang2018a}; however, care must be taken to select enough appropriate points to simultaneously capture the shape of the function as well as correlations between devices. 

\subsection{Physics-informed kernel\label{sec:gp:kernelphysics}}


In order to introduce physics relations and correlations into the kernel in the absence of data, we suggest a method to calculate the kernel directly from a physics simulation. As such, there is no need to assume a specific parameterized covariance kernel and solve an optimization problem for the hyper-parameters of the kernel.


The kernel describes the target function's expected behavior. Since a good regulator of a system is a model of that system  \cite{good_regulator}, the best estimator of the target's covariance is the target function itself. However, the target may be costly to evaluate, so a faster evaluating physical model or simulation approximating the system to optimize may be used instead. 
Rather than calculate the covariance of the approximating system directly, we expand the simulation $f(x)$ about the peak or optimum point $x_0$ with a Gaussian: (Eq. \ref{basis_fcn}) and then use this to calculate the associated RBF covariance function. This can be done quickly by evaluating the Hessian $H_{i,j}=\partial_{x_i}\partial_{x_j}f(x)|_{x=x_0}$ of the simulation. 

\begin{equation}
    \phi(x) = f(x_0) \exp(-\frac{1}{2} (x-x_0)^T \frac{-H}{f(x_0)} (x-x_0))
    \label{basis_fcn}
\end{equation}

Since this expansion is Gaussian, application of Eq.~\ref{eq:kernel_from_basis} yields an RBF covariance function $k(x,x')=\sigma_f^2 \exp(-(x-x')^T\Sigma(x-x')/2)$ with precision matrix half that of the expansion above $\Sigma=-H/2f(x_0)$. 
The resulting covariance function then approximates the covariance of the target system. If the system converges asymptotically to a constant value as in the physical system controlled below, we subtract off this asymptotic constant, and replace $f(x_0)$ in the equations above with $f(x_0)-f(\infty)$. The function value $f(\infty)$ then becomes the Gaussian process prior mean, or we can use the simulation itself or an approximation as an explicit prior.

\section{Online demonstration on SPEAR3 storage ring}
In this section, we experimentally demonstrate the effectiveness of the physics-informed kernel with an example from particle accelerator physics. We show that when physics-informed correlations \cite{Duris2019} and length scales are accounted for, online optimization of an objective function is faster.

\subsection{Minimization of storage ring vertical emittance}


SPEAR3~\cite{SPEAR3}, a successor of the Stanford Positron Electron Asymmetric Ring, is a 3-GeV, high-brightness third generation storage ring operating with high reliability and low emittance. It runs with 500 mA in top-off mode, during which the beam current is kept constant with injection of electrons into the ring every five minutes.

Emittance is a measure of the phase space area occupied by a particle beam in one degree of freedom of beam motion. 
A small emittance in a storage ring light source is preferred as it results in high photon beam brightness. 
In an ideal electron storage ring, the vertical emittance is nearly zero. However, there are various sources of errors that give rise to 
a finite vertical emittance in an operating electron storage: for example, vertical orbit distortion and linear betatron coupling between the horizontal and vertical planes. Those error sources contributing to the vertical emittance can be compensated with skew
quadrupole magnets. In SPEAR3, there are 13 free skew quadrupoles for vertical emittance control. Usually the skew quadrupole setting for vertical emittance correction is obtained by fitting the orbit response matrix data \cite{Safranek1997}, in particular, the off-diagonal elements.

In this experiment our goal is to minimize vertical emittance with skew quadrupoles. For most third generation light sources, the beam loss is dominated by Touschek scattering. In such case, the square root of the vertical emittance is inversely proportional to the beam loss rate. Thus, minimizing vertical emittance is equivalent to maximizing beam loss rate with respect to skew quadrupoles \cite{Huang2013a}.


\subsection{Data-informed kernel \label{sec:3GPdata}}

Motivated by the observation that the negative beam loss rate response to skew quadrupole magnets looks Gaussian, we chose the radial basis function (RBF) kernel with a Gaussian noise kernel, i.e. $k(x_i,x_j) = \sigma_f^2 \exp(-\frac{1}{2}(x_i - x_j)^{\rm T} \Sigma (x_i -  x_j)) +  \sigma_{n}^2 \delta(x_i - x_j)$.
This kernel depends on several hyper-parameters: $\sigma_f^2$ is the covariance function amplitude,  $\Sigma$ is the precision matrix, $\sigma_{n}^2$ is the noise variance, which models the variance of the prediction at a sampled point, and $\delta$ is the Dirac delta function. 
In order to find a set of good hyper-parameters $\theta =\{ \sigma_{f}^2, \sigma_{n}^2, \Sigma\}$, we maximize the log-marginal likelihood expression for archived data of slice scans about the peak in each dimension. 

\subsection{Physics-informed kernel\label{sec:kernelphysics}}

As a first step in the full treatment of constructing basis functions from simulations, we 
approximated the SPEAR3 storage ring simulation with a Gaussian as described in Section~\ref{sec:gp:kernelphysics}. We subtracted off the asymptotic beam loss rate of 0.57~mA/min found in the simulation, and then calculated the precision matrix from the Hessian of the simulation.
%
%
%
We calculated the Hessian in two ways.
First, we interpolated the results of densely sampled simulation data with a neural network regressor, and calculate the hessian. Second, we calculated the 
hessian at the maximum beam loss rate $(L)$ point $H_{i,j}=\partial_{x_i} \partial_{x_j}L$. The two methods are in a relative good agreement.

While the full precision matrix of the kernel, which takes into account both lengthscales and correlations, was calculated from physics simulations, the amplitude and noise parameters of the kernel were calculated from machine data, similar to \autoref{sec:3GPdata}.



\subsection{Experiment and Results}

The loss rate (mA/min) is measured by monitoring the beam current drop in a fixed time interval. The skew quadrupole parameter current range is $(-20,20)~{\rm A}$.
The noise in the beam loss rate data is mostly from the uncertainty in the beam current measurement, and the rms noise of the loss rate as measured by the standard deviation residual deviations from expectation over 300 seconds was 0.04 mA/min.
The quads were set to zero before scanning each time and this reduced the beam loss rate to 0.45 mA/min. The loss rate is evaluated by computing the change in the beam current loss rate over one second. Then we waited two seconds to let the devices settle in the next point.

We tested the GP optimizer with an upper confidence bound (UCB) acquisition function \cite{ucb}, which is constructed from the GP prediction mean and variance. The GP optimizer is initialized with a kernel and first observed point. Two different kernels were tested, one derived form data (\autoref{sec:3GPdata}), and the other from physics (\autoref{sec:kernelphysics}). Moreover, we tested the Nelder Mead simplex optimization method \cite{simplex}, which is a standard algorithm routinely used to tune particle accelerator systems \cite{tomin:ocelot}.
%

\begin{figure}[htpb!]{}
\centering
 \includegraphics[width= 0.45\textwidth]{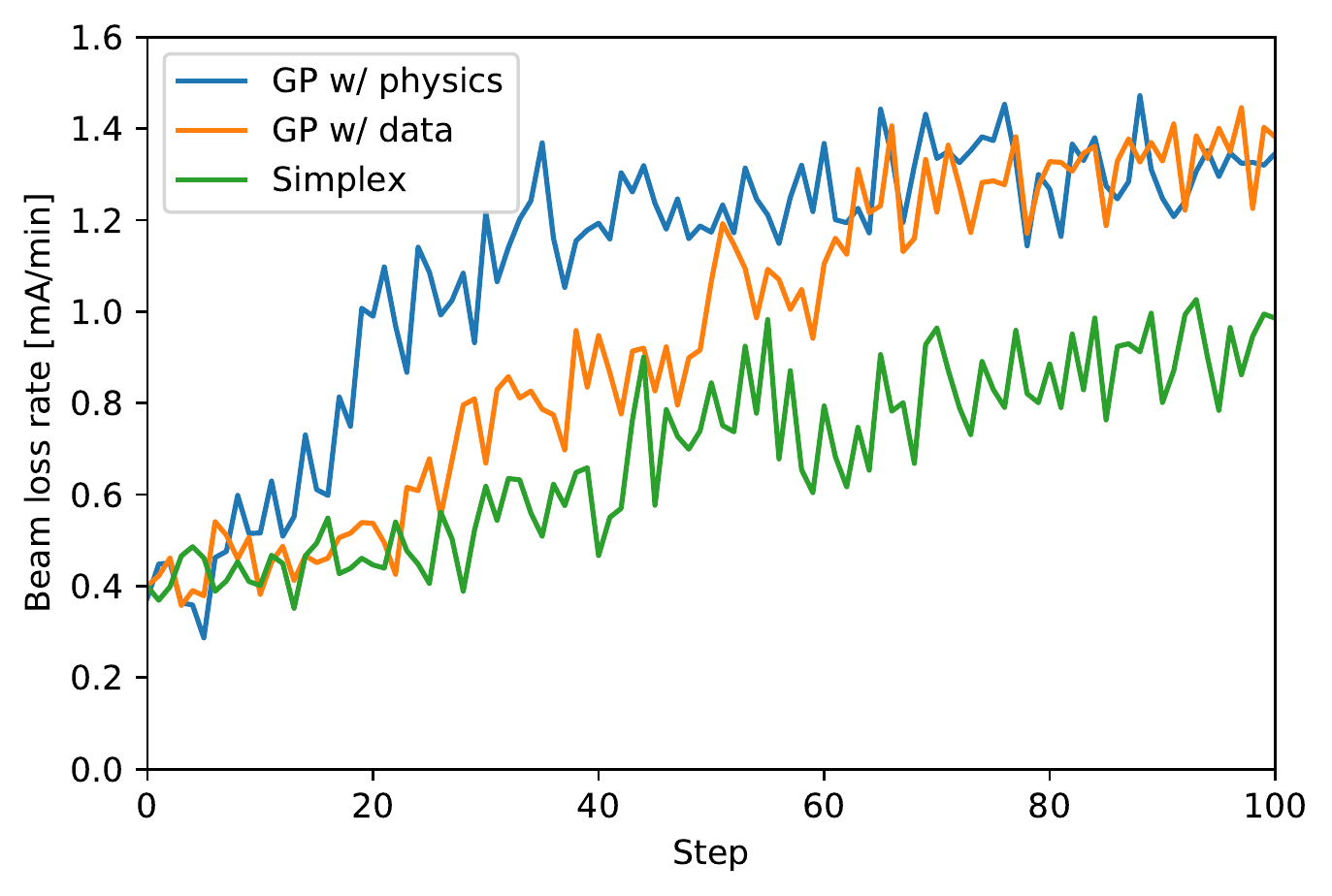}
  \caption{Comparison of optimization of beam loss rate with respect to 13 skew quadrupole magnets for Gaussian process optimization with physics-informed kernel from physics (blue) vs Gaussian process with data-informed kernel (orange) vs Nelder-Mead simplex optimizer (green). Each step corresponds to approximately 3 seconds of beam time. The GP with physics-informed kernel converge faster to a higher optimum as compared with the other two mehods.}
  \label{fig:comparison}
\end{figure}

%
%
Figure \ref{fig:comparison} shows results from online optimization of the beam loss rate simultaneously on 13 skew quadrupole magnets. 
The Gaussian process optimizer with physics-informed kernel (blue curve) reaches a higher optimum (1.4 mA/min) than either the Gaussian process optimizer with kernel from archive data (orange curve) or the Nelder-Mead simplex algorithm (green curve). Furthermore, the GP with the physics-informed kernel converges faster to this optimum; 80 to 100 steps (3 to 5 min) per scan. 

\section{Summary}


 In this work we showed an alternative method to Bayesian optimization that pivots from a pure historical data approach, which obviously fails for new machines and configurations, to one which relies on physics models. The approach may be more robust and flexible than the pure data-driven approach. 
This method could be applicable to automatic tuning of complex systems with complicated setups. We demonstrated an online optimization of electron beam loss rate with respect to 13 skew quadruples on the SPEAR3 storage ring. We experimentally showed that Gaussian process optimizer with physics-informed kernel from simulation is faster than both uncorrelated kernel deduced from archive data, and from the standard simplex optimizer.  
Quickly calculating kernels from an online model is attractive for free-electron laser light sources which change configurations multiple times per day.
%
We anticipate that the ability to insert physics into GPs will turn them into an attractive method for practitioners in various domains, and may have wide applications in science.

\subsubsection*{Acknowledgments}
The authors are grateful to the SPEAR3 operators and engineers for their help with live tests on the storage ring.
This work was supported by the Department of Energy, Laboratory Directed Research and Development program at SLAC National Accelerator Laboratory, under contract DE-AC02-76SF00515, and by Office of Advanced Scientific Computing Research under FWP 2018-SLAC-100469ASCR.


\bibliographystyle{unsrt}
\bibliography{references}

\begin{thebibliography}{10}

\bibitem{rw:gpml_book}
Songthip~T Ounpraseuth.
\newblock {\em {Gaussian Processes for Machine Learning}}, volume 103.
\newblock MIT Press, 2008.

\bibitem{w:gpml_book_ch4}
Carl~Edward Rasmussen and Christopher K.~I. Williams.
\newblock {Gaussian Processes for Machine Learning: CH4. Covariance Function}.
\newblock In {\em Gaussian processes for machine learning}, volume~88. 2006.

\bibitem{duvenaud:kernel_composition}
David Duvenaud, James~Robert Lloyd, Roger Grosse, Joshua~B. Tenenbaum, and
  Zoubin Ghahramani.
\newblock {Structure Discovery in Nonparametric Regression through
  Compositional Kernel Search}.
\newblock Technical report, 2013.

\bibitem{Wilson2013}
Andrew~Gordon Wilson and Ryan~Prescott Adams.
\newblock {Gaussian process kernels for pattern discovery and extrapolation}.
\newblock Technical Report PART 3, 2013.

\bibitem{Bozson2018}
Adam Bozson, Glen Cowan, and Francesco Span{\`{o}}.
\newblock {Unfolding with Gaussian Processes}.
\newblock Technical report, 2018.

\bibitem{Frate2017}
Meghan Frate, Kyle Cranmer, Saarik Kalia, Alexander Vandenberg-Rodes, and
  Daniel Whiteson.
\newblock {Modeling Smooth Backgrounds and Generic Localized Signals with
  Gaussian Processes}.
\newblock Technical report, 2017.

\bibitem{DKL}
Andrew~Gordon Wilson, Zhiting Hu, Ruslan Salakhutdinov, and Eric~P. Xing.
\newblock {Stochastic variational deep kernel learning}.
\newblock In {\em Advances in Neural Information Processing Systems}, pages
  2594--2602, 2016.

\bibitem{manifoldgp}
Roberto Calandra, Jan Peters, Carl~Edward Rasmussen, and Marc~Peter Deisenroth.
\newblock {Manifold Gaussian Processes for regression}.
\newblock Technical report, 2016.

\bibitem{Deep-GP}
Andreas~C Damianou and Neil~D Lawrence.
\newblock {Deep Gaussian Processes, Damianou {\&} Lawrence}.
\newblock Technical report, 2013.

\bibitem{Duris2019}
Joseph Duris, Dylan Kennedy, Adi Hanuka, Jane Shtalenkova, Auralee Edelen, Adam
  Egger, Tyler Cope, and Daniel Ratner.
\newblock {Bayesian optimization of a free-electron laser}.
\newblock {\em arXiv: 1909.05963}, 2019.

\bibitem{Yang2018a}
Xiu Yang, Guzel Tartakovsky, and Alexandre Tartakovsky.
\newblock {Physics-Informed Kriging: A Physics-Informed Gaussian Process
  Regression Method for Data-Model Convergence}.
\newblock Technical report, 2018.

\bibitem{good_regulator}
Roger~C. Conant and W.~Ross~Ashby.
\newblock {Every good regulator of a system must be a model of that system}.
\newblock Technical Report~2, 1970.

\bibitem{SPEAR3}
R~Hettel.
\newblock {The Completion of SPEAR 3}.
\newblock In {\em 9th European Particle Accelerator Conference}, 2004.

\bibitem{Safranek1997}
J.~Safranek.
\newblock {Experimental determination of storage ring optics using orbit
  response measurements}.
\newblock {\em Nuclear Instruments and Methods in Physics Research, Section A:
  Accelerators, Spectrometers, Detectors and Associated Equipment},
  388(1-2):27--36, 3 1997.

\bibitem{Huang2013a}
Xiaobiao Huang, Jeff Corbett, James Safranek, and Juhao Wu.
\newblock {An algorithm for online optimization of accelerators}.
\newblock Technical report, 2013.

\bibitem{ucb}
Niranjan Srinivas, Andreas Krause, Sham~M. Kakade, and Matthias~W. Seeger.
\newblock {Gaussian Process Optimization in the Bandit Setting: No Regret and
  Experimental Design}.
\newblock {\em Proceedings of the 27th International Conference on Machine
  Learning (ICML 2010)}, pages 1015--1022, 2010.

\bibitem{simplex}
J.~A. Nelder and R.~Mead.
\newblock {A Simplex Method for Function Minimization}.
\newblock {\em The Computer Journal}, 7(4):308--313, 1 1965.

\bibitem{tomin:ocelot}
S.~Tomin, G.~Geloni, I.~Agapov, I.~Zagorodnov, Ye. Fomin, Yu. Krylov,
  A.~Valintinov, W.~Colocho, T.M. Cope, A.~Egger, and D.~Ratner.
\newblock {Progress in automatic software-based optimization of accelerator
  performance}.
\newblock In {\em IPAC 2016 - Proceedings of the 7th International Particle
  Accelerator Conference}, 2016.

\end{thebibliography}

\end{document}